\documentclass[twocolumn,amssymb]{revtex4} 
\usepackage{epsfig,amssymb,amsmath,graphicx,subfigure,hyperref  }  
\usepackage{color}

\begin{document}

\title{Propagating stress-pulses and wiggling transition revealed in
string dynamics} 
\author{Zhenwei Yao} 
\email{zyao@sjtu.edu.cn}
\affiliation{School of Physics and Astronomy, and Institute of Natural
Sciences, Shanghai Jiao Tong University, Shanghai 200240 China}  
\begin{abstract} 
Understanding string dynamics yields insights into the intricate
dynamic behaviors of
various filamentary thin structures in nature and industry covering multiple length
scales. In this work, we
investigate the planar dynamics of a flexible string where one end is free and
the other end is subject to transverse and longitudinal motions. Under transverse
harmonic motion, we reveal the propagating pulse structure in the stress profile
over the string, and analyze its role in bringing the system into a chaotic
state. For a string where one end is under longitudinal uniform acceleration, we identify the wiggling
transition, derive the analytical wiggling solution from the string equations, and
present the phase diagram. 
\end{abstract}
\maketitle

\section{Introduction}

An inextensible flexible string is the backbone of many complicated
quasi-one-dimensional thin objects, and represents one of the simplest
organization of matter~\cite{whittaker1988treatise, Audoly2010b, o2017modeling}.
Interest in inextensible flexible strings can be traced back to the beginnings
of the calculus~\cite{reeken1977equation}.  When in motion, a flexible string can
exhibit a number of counterintuitive dynamic behaviors, ranging from the
acceleration of a string when striking a
table~\cite{grewal2011chain,grewal2011erratum,2017arXiv171205778C}, the
formation of the chain fountain structure~\cite{biggins2014understanding,
biggins2014growth,virga2014dissipative,pantaleone2017quantitative}, to the
spontaneous rise-up and lift-off of a pulled string in the
plane~\cite{hanna2011instability, hanna2012slack} and on a
pulley~\cite{cambou2012unwrapping, brun2016surprising}.  Understanding the
intricate dynamics of the filamentary string structure is important as
they are ubiquitous in nature and industry covering length scales of several
orders of
magnitude~\cite{costello1997theory,neukirch2004extracting,carter2009submarine,o2017modeling}.  Much has been learnt about the string
dynamics by analyzing the equations of motion of the
string~\cite{broer1970dynamics, edwards1972dynamics,
reeken1977equation,schagerl2002propagation,brun2016surprising,csengul2017generalized}.
However, the analytical solution to the coupled, nonlinear string equations is only
limited to some special cases~\cite{fetter2003theoretical,
hanna2012end,hanna2012slack}. Particle simulation based on the spring-bead
model has proven to be a powerful tool to study the dynamic states of the
string~\cite{milchev1996static, rapaport2004art, hanna2011instability,
o2017modeling}.

The goal of this work is to explore the planar dynamics of the string where one
end is free and the other end is in transverse and longitudinal motion,
respectively. This model system provides the opportunity to clarify a host of
questions with broader implications, such as: How will the motion at one end of
the string propagate to the other end? Will any dynamic instability occur in the
string? What kinds of featured dynamic states will emerge? To address these
questions, we resort to the combination of particle simulations and theoretical
analysis of the string equations. The main results of this work are presented
below. When shaking the string at one end in harmonic motion, we find that the
propagation of stress is realized by the oscillating stress-pulse structure
across the string. The back-and-forth movement of the stress-pulse induces more
pulses and ultimately leads the whole string to a chaotic state. For a traveling
string in uniform acceleration, we find a new dynamic state of the string in which
it starts to wiggle and deviate from the straight shape after finite
duration. We derive an
analytical wiggling solution from the string equations which can substantiate
the numerical observation. We further characterize the wiggling transition,
and present the phase diagram.

\begin{figure*}[thb]  
\centering 
\includegraphics[width=7in]{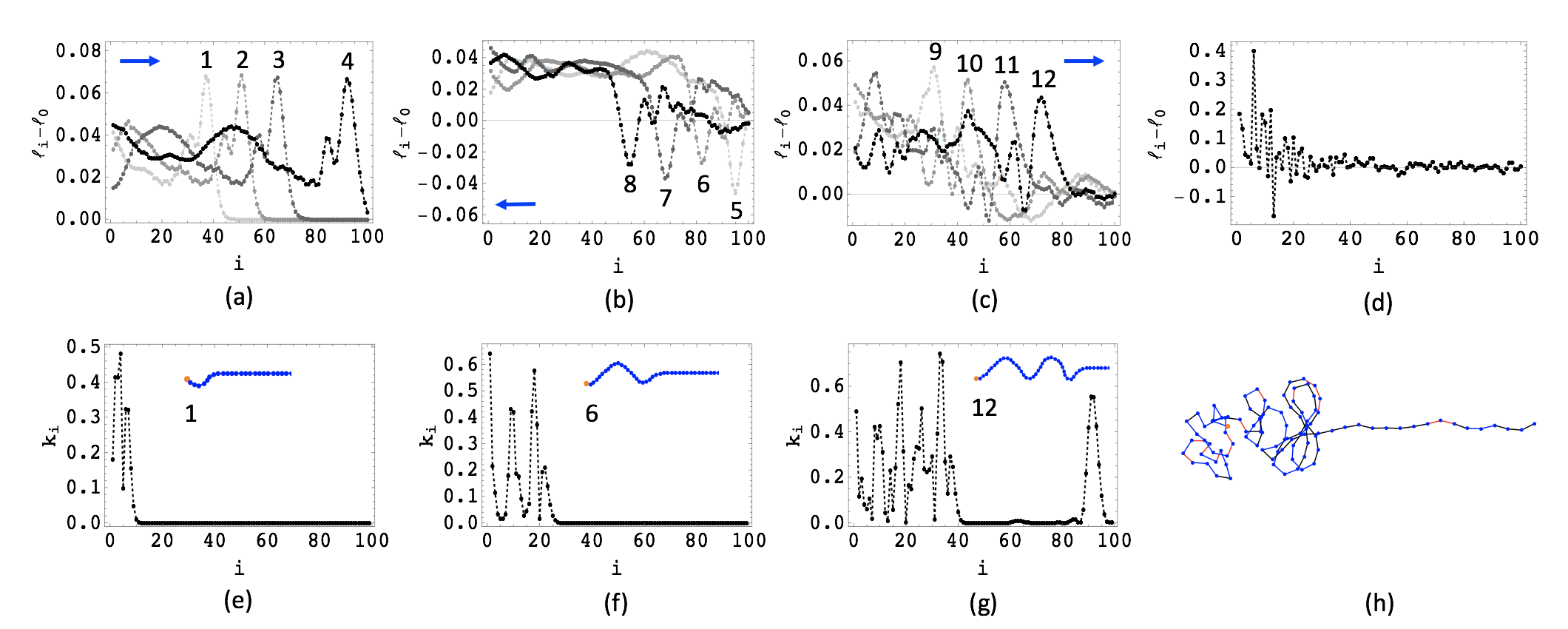}
\caption{The propagating stress-pulse structure in the string when one end is under transverse
  harmonic oscillation with amplitude $A=2.5\ell_0$ and period $T=100\tau_0$.
  (a)-(d) show the distribution of the bond length. The labeled numbers indicate the temporal sequence: from 1 to 12, $t/\tau_0=42$, 56, 70, 98, 112, 126, 140, 154,
  182, 196, 210, 224. $t/\tau_0=700$ in the rightmost figures. (e)-(g)
  show the distribution of curvature $\kappa$ in the typical conformations of the
  string. Note that only a small part of the straight segment is shown in the insets
for visual convenience.  $A_{\textrm
  {noise}}=10^{-5}\ell_0$. }
\label{impulse}
\end{figure*}

\section{Model and Method}

An inextensible flexible string can be modeled by the geometric curve $\vec{X}(s,
t)$, where $s$ is the natural parameter of the curve and $t$ is the time. The
inextensibility condition is $\partial_s \vec{X}(s, t) \cdot \partial_s
\vec{X}(s, t) = 1$. The dynamics of the flexible, inextensible string with uniform mass density $\mu$
is governed by the following equation of motion~\cite{broer1970dynamics, reeken1977equation} 
\begin{eqnarray}
\mu \partial^2_t \vec{X}(s, t) = \partial_s [\sigma(s,t) \partial_s \vec{X}(s,
      t)],\label{eom}
\end{eqnarray}
where the stress $\sigma$ arises as a Lagrangian
parameter to keep
neighbouring parts of the string at fixed distance~\cite{reeken1977equation}. 
By projecting eq.(\ref{eom}) along
the tangent and normal vectors, we obtain the following string
equations~\cite{hanna2012end}:
\begin{eqnarray}
\sigma \kappa^2 -\partial_s^2 \sigma = \mu \partial_t \hat{t} \cdot \partial_t
\hat{t}, \label{eom1} \\
2 \kappa \partial_s \sigma + \sigma \partial_s \kappa = \mu \partial_t^2 \hat{t}
\cdot \hat{n}, \label{eom2}
\end{eqnarray}
where $(\hat{t}, \hat{n})$ is the dyad of unit tangent and normal vectors on a planar curve, and $\kappa$ is
the curvature. It is a challenge to analytically solve the coupled
nonlinear differential equations~\cite{reeken1977equation}. Furthermore, due to its flexibility, 
the string may exhibit shapes that are beyond the functional space of
$\vec{X}(s, t) \in C^2(Q_T)$  and $\sigma(s, t)
\in C^1(Q_T)$, where $Q_T=J_s \times J_t$, $s \in J_s=[0,
L]$, and $t \in J_t = [0, T]$~\cite{reeken1977equation}.
The above string equations lay the
foundation for the theoretical analysis of relevant simulation results.

In our simulations, the string is modeled
by $N+1$ massive beads connected by high stiffness linear springs lying on the plane. The
balance length of each spring is $\ell_0\equiv 1$. The mass of each bead is
$m_0\equiv 1$. In the initial state, the string is free of stress, and lies
along the x-axis. 
The position of each bead is subject to a 
 small quantity of
 noise $\delta
\vec{x}$ whose x- and y-components conform to the uniform distribution in the
interval $[-A_{\textrm{noise}},  A_{\textrm {noise}}]$. The introduction of
this noise is to trigger a possible instability of a string in longitudinal
motion, and also reflects the small fluctuation of the string under various noise
sources in reality.  We implement the
Verlet integration to construct the trajectory of each bead in the discretized
string~\cite{rapaport2004art}. We work in the
regime of highly inextensible string with large $k_0$. Specifically, $\tilde{T}
\equiv T/\tau_0 = T \sqrt{k_0}/\sqrt{m_0} >>1$ in the transverse harmonic
oscillation of period $T$, and $\tilde{a} \equiv a/a_0 = a m_0/(\ell_0 k_0)<<1$
in the case of longitudinal uniform acceleration.  $\tau_0=\sqrt{m_0/k_0}=1$.
$a_0=\ell_0/\tau_0^2=1$.

\section{Transverse harmonic oscillation} In this section, we present the main results about the
planar dynamics of the string when one end is under transverse harmonic
oscillation. The motion of the shaking
end (labeled as $i=0$) is $\{x_0(t)=0, y_0(t) =
A\sin\left(2\pi t/T \right)\}$.

We tune the amplitude of the noise in the position of each bead to be
a very small fraction of the balance length $\ell_0$ of the spring, and work in the
regime of large $T$ (i.e., highly inextensible string). Simulations with
varying shaking amplitude $A$ from $\ell_0$ to $5\ell_0$ show that in general
the harmonic motion at the head of the string can propagate in the form of a
cosine-like wave by only a few wavelengths. In fig.~\ref{impulse}(e)-(g), we
present the typical case of $A=2.5\ell_0$ and $T=100\tau_0$. The entire string
consists of the straight and the
wavy segments. The horizontal orientation of the tangent vector at the
connection of the straight and the wavy parts of the string seems crucial for
maintaining the straight segment of the string. Continuously shaking the string finally leads to
the chaotic state as shown in fig.~\ref{impulse}(h), which is characterized by
the large deformation of the waves in the head part and the growing
transverse fluctuation in the remaining part of the string.

The formation of the wave structure near the shaking end reduces the
longitudinal length of the string due to the rigidity of the spring. The
realization of the geometric shrinking of the string relies on the propagation
of stress. The question of how the stress propagates across the string naturally
arises. In the following, we analyze the evolution of the stress profile over
the string in this process. The results are summarized in fig.~\ref{impulse}.

Figures~\ref{impulse}(a)-(d) shows the variation of the stress distribution over
the string as it evolves towards the chaotic state. The shaking bead is at
$i=0$. The labeled numbers at the peaks indicate the temporal sequence, some of
which correspond to the labeled shapes in fig.~\ref{impulse}(e)-(g).
Simulations reveal the peak structure in the stress profile. It indicates that
the stress propagates in the manner of pulses.  The peak structures in the
stress profile are named as stress-pulses. The stress-pulse region, where
$l-l_0>0$, is stretched much more than the remaining part of the string.  From
fig.~\ref{impulse}(a), we see that the region from the location of the
stress-pulse to the free end is free of stress. The stress-pulse sharply
separates the stretched and the stress-free regions. Here, we
emphasize that the word ``pulse" specifically refers to the peak structure in the
stress profile in fig.~\ref{impulse}(a)-(d), but not the wave structure in the
string shape as shown in fig.~\ref{impulse}(e)-(g). Simulations show the
steady propagation of the stress-pulse across the string in a rate that is
much faster than the propagation of the wave in the string shape, as shown in
fig.~\ref{impulse}(a). From fig.~\ref{impulse}(a), we obtain the value
of the pulse speed to be the characteristic speed of the string $\ell_0/\tau_0$,
which is proportional to $\sqrt{k_0}$. Further simulations for the case of
$T/\tau_0=1000$, which is ten times the value for $T/\tau_0$ in
fig.~\ref{impulse}, confirm that the pulse speed is the same as that in
fig.~\ref{impulse}. Therefore, the pulse propagates 
infinitely fast in the inextensible limit.

Figure~\ref{impulse}(b) shows that when the stress-pulse labeled $4$ in
fig.~\ref{impulse}(a) reaches the free end
of the string, it is bounced back, becoming the pulse labeled $5$. Remarkably,
the stress-pulse is inverted in this process. In other words, the pulse region that is
originally stretched becomes compressed. Consequently, the stress distribution over the
string is divided into a number of compressed and stretched regions; the shaking
end is always stretched. In contrast, for a pulse whose
dynamics is governed by the wave equation, no inversion occurs when reflecting
off a free end of the medium~\cite{Arfken1999}. Here, the behavior of the
stress-pulse in the highly inextensible string system is governed by the
coupled string equations in eq.(\ref{eom1}) and (\ref{eom2}) rather than the law
of the wave equation.  Figure~\ref{impulse}(b) shows that the negative
stress-pulse continues propagating towards the shaking end. It is finally
reflected back, and becomes inverted again [see the pulse labeled $9$ in
fig.~\ref{impulse}(c)]. It is of interest to note that propagation and reflection of small waves
along a hanging chain subject to an initial velocity have been
studied~\cite{schagerl2002propagation} and an interesting pattern of kicks at the free end has been
revealed~\cite{bailey2000motion}.

\begin{figure}[t]  
\centering 
\subfigure[]{
\includegraphics[width=1.6in]{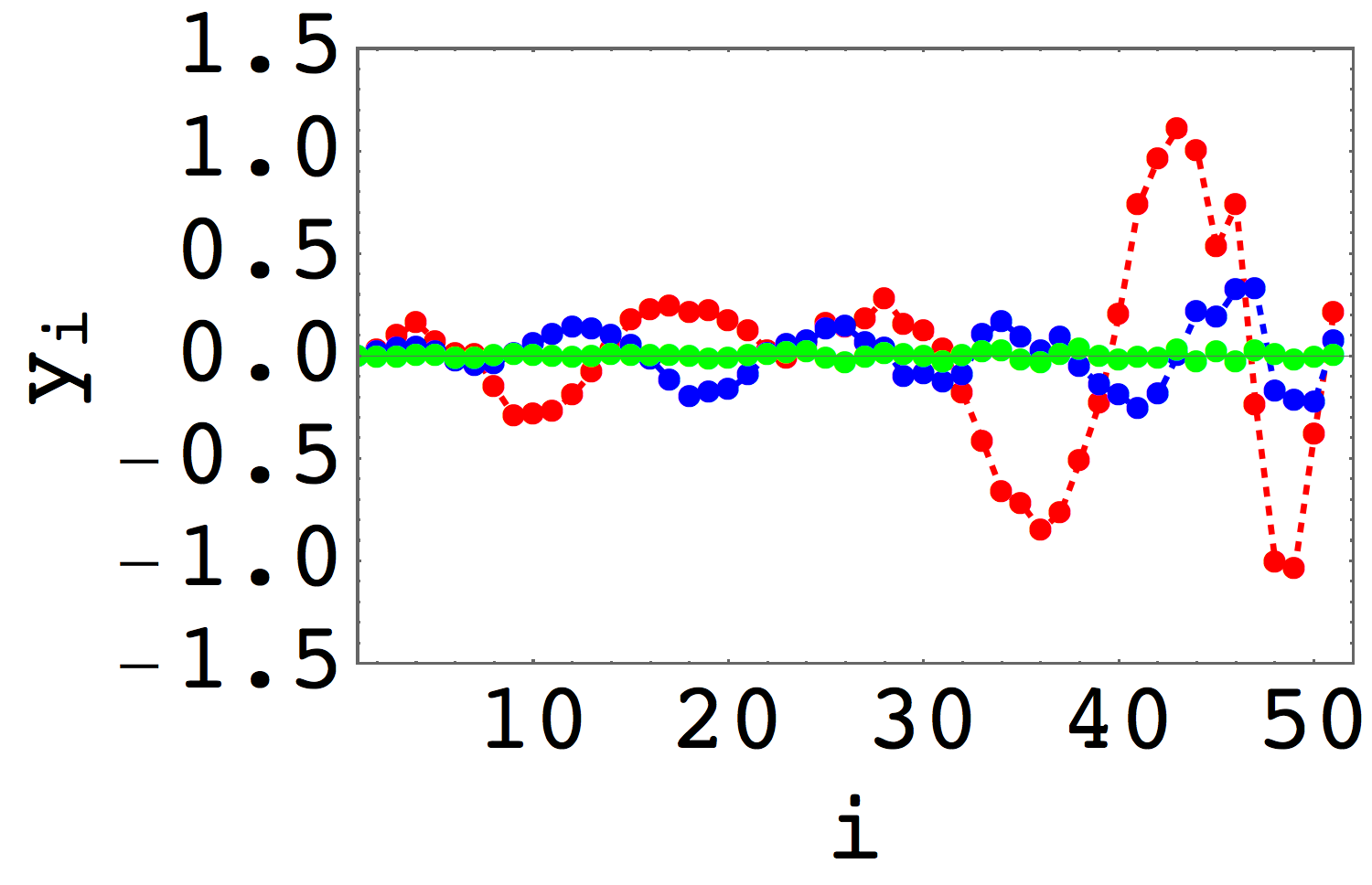}}  
\hspace{-0.0in} 
\subfigure[]{
\includegraphics[width=1.6in]{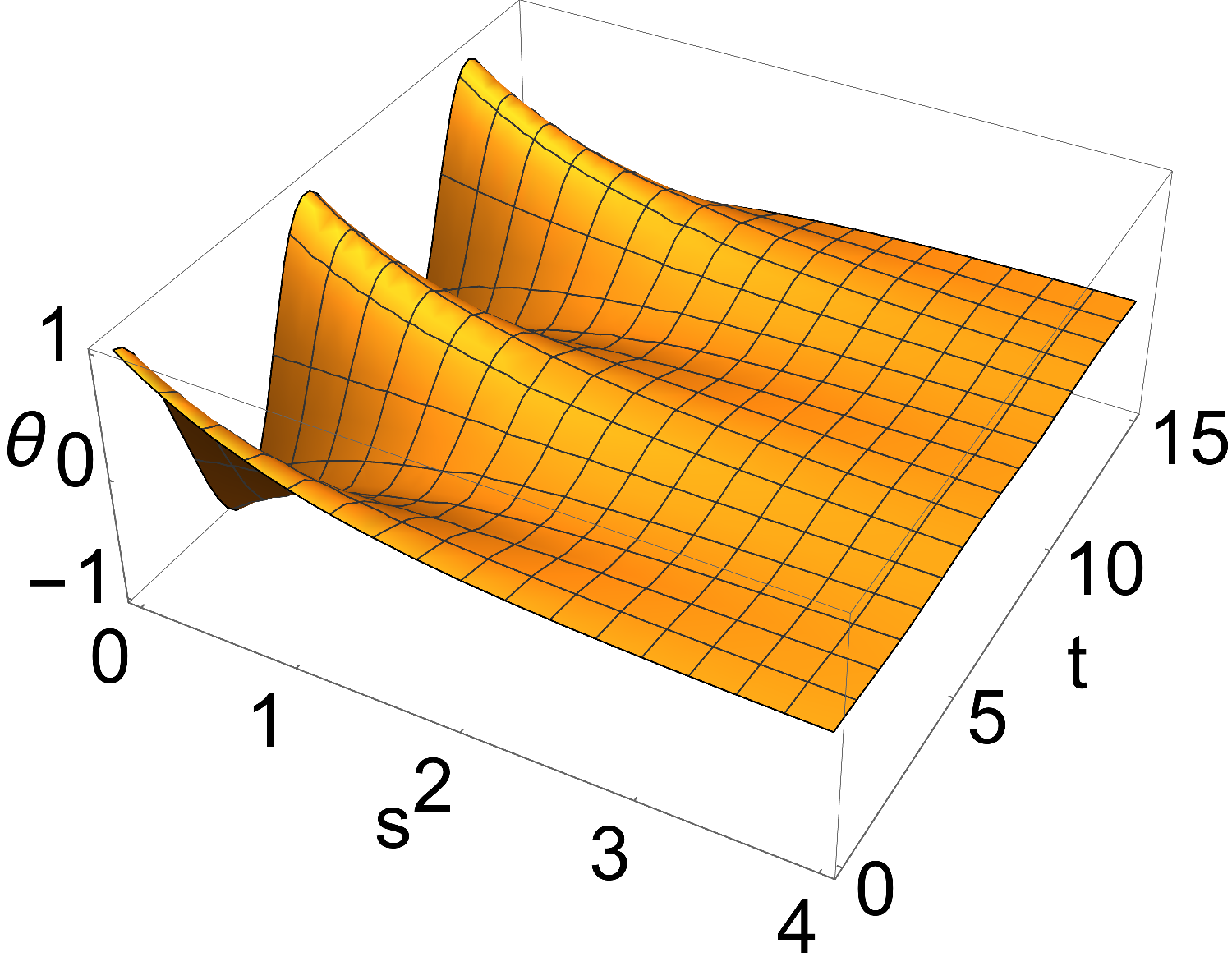}}
\caption{Wiggling of the string when one end is under longitudinal uniform
  acceleration. (a) $y_i$ is the transverse displacement of the
    beads. $a/a_0=10^{-4}$ (green), $10^{-3}$
    (blue), and $10^{-2}$ (red). $t=1000\tau_0$. $A_{\textrm
      {noise}}=10^{-3}\ell_0$. (b) Plot of the wiggling solution
    derived from the string equations. $\theta(s,t) = \theta_0 \exp(-qs)\cos(wt)$,
  where  $q=1$, $w=1$, and $\theta_0=1$.}
\label{acc}
\end{figure}

With the back-and-forth movement of the stress-pulse, we numerically observe the
continuous retreat of the free end towards the shaking end along the axis of the
string. Furthermore, the oscillation of the stress-pulse
across the string induces more pulses as shown in fig.~\ref{impulse}(c).
Repeating this process ultimately destroys the wavy shape near the shaking end,
and the shape of the string becomes chaotic as shown in fig.~\ref{impulse}(h).
The transition to the chaotic state is also reflected in the stress profile.
From fig.~\ref{impulse}(d), we see that the stress is highly concentrated in the
chaotic segment
of the string, and the stress level at the straight segment is significantly reduced. 
The screening of the stress by
the highly curved segment in
the chaotic string can be rationalized by the first term in eq.(\ref{eom1}).
Equation (\ref{eom1}) is recognized as the
screened Poisson's equation $(\frac{d^2}{dx^2}-k^2)\psi =
f(x)$ for constant $\kappa$; the source
term is the temporally varying tangent vector~\cite{Arfken1999}. The corresponding Green's
function is $G(x_1,x_2)=\frac{1}{2k}e^{-k|x_1-x_2|}$ under the boundary condition
that the Green's function vanishes for $x\rightarrow \pm \infty$. Therefore,
the effect of curvature in the string is to screen the stress.

\section{Longitudinal uniform acceleration} 
We proceed to discuss the planar dynamics of the string under
longitudinal uniform acceleration based on simulations and theoretical analysis. The head bead of the string is pulled and
maintained in uniform acceleration
along the x-axis: $x_0(t)=\frac{1}{2}a t^2$. For a straight string in
longitudinal uniform acceleration, the string equations show that the stress is linear with $s$, increasing from zero to $\mu a L$
from the tail ($s=0$) to the head ($s=L$) of the string. However, simulations
reveal that the traveling string will suddenly deviate from the
straight shape after finite duration, and the shape fluctuation persists thereafter. We name such a
dynamic transition as the wiggling transition. 

\begin{figure}[h]  
\centering 
\subfigure[]{
\includegraphics[width=1.6in]{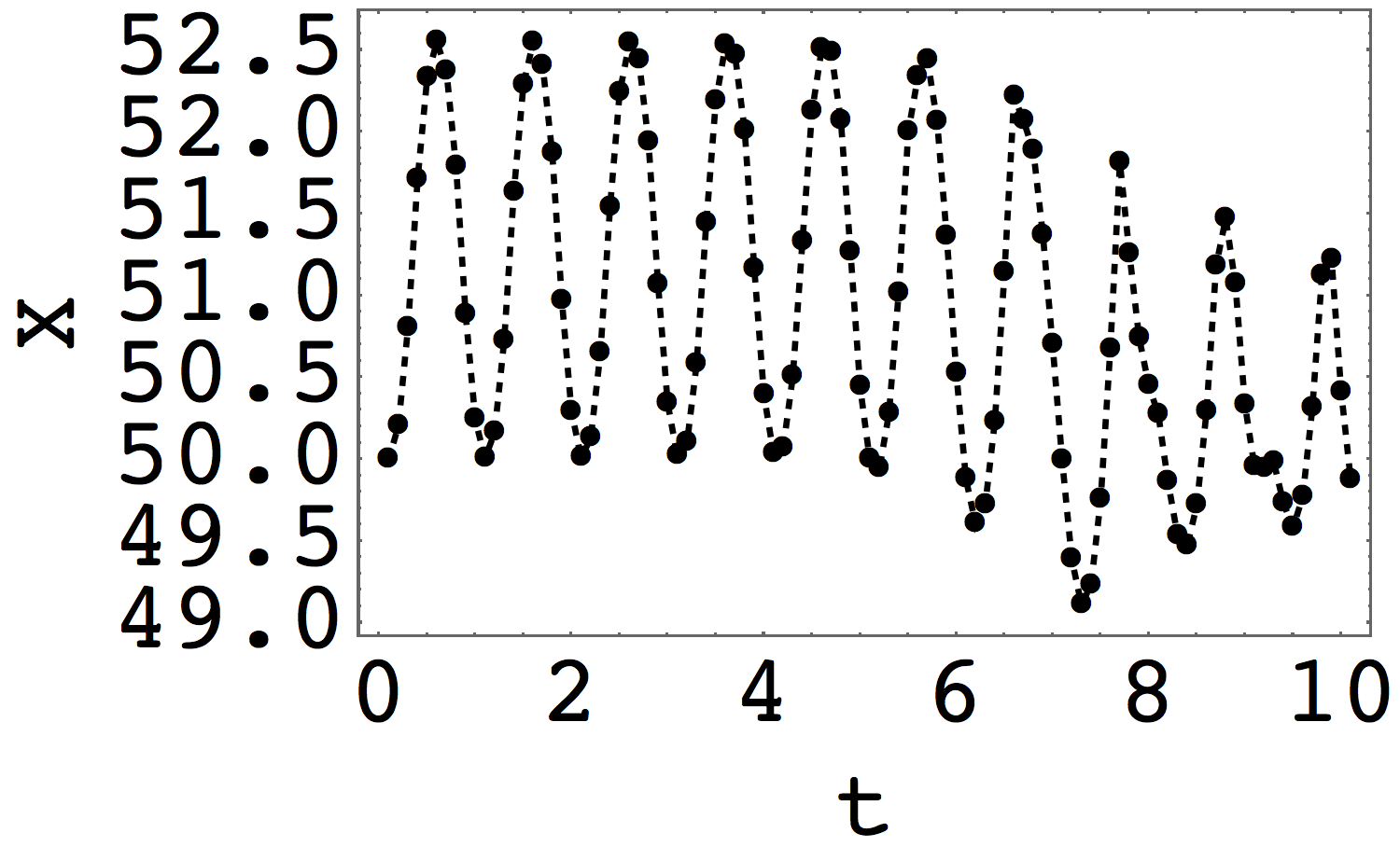}}
\subfigure[]{
\includegraphics[width=1.66in]{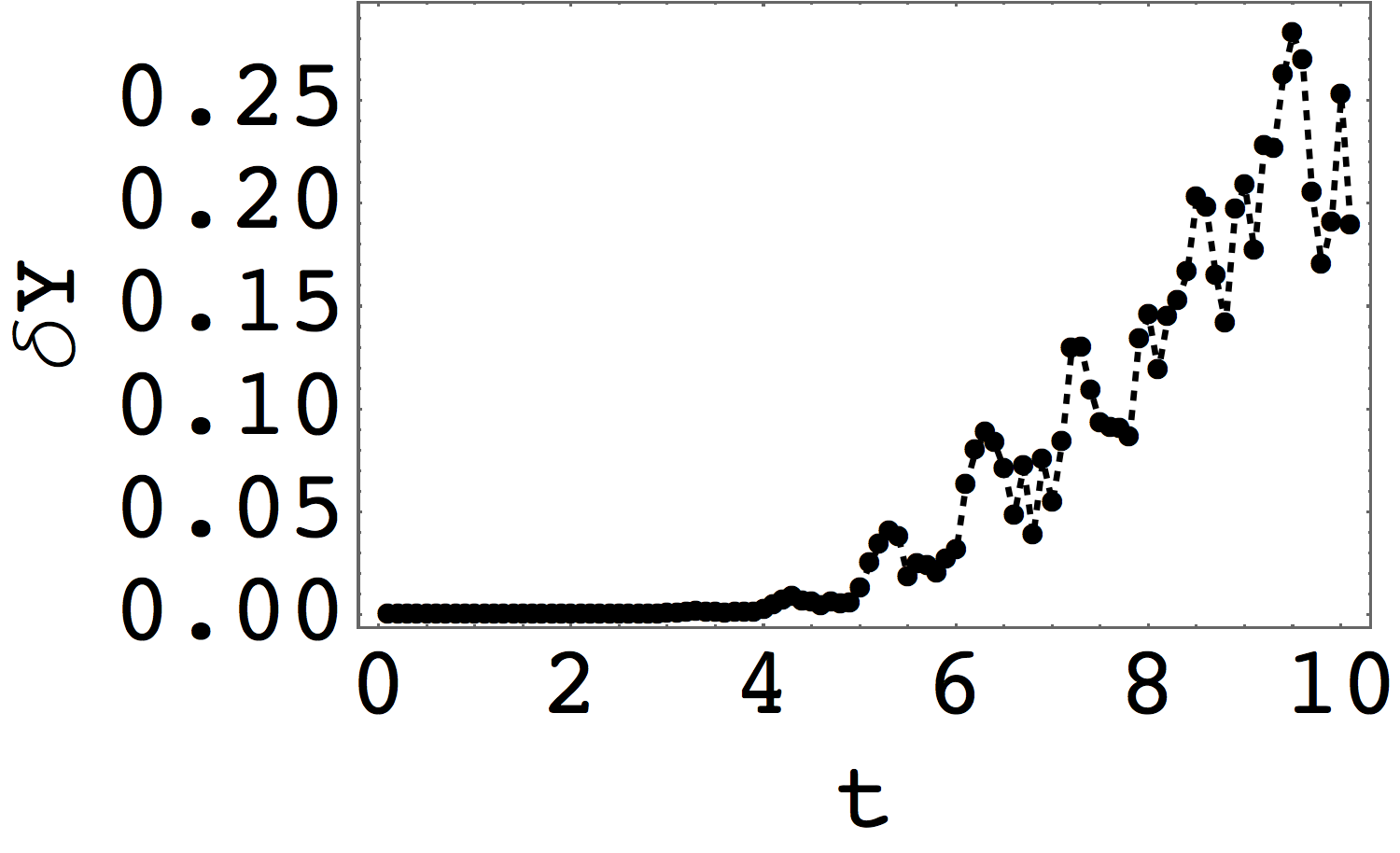}}  
\caption{Characterization of the wiggling phenomenon of the traveling string in
  uniform acceleration. (a) and (b) Plots of the longitudinal length $X$ and the
  averaged transverse displacement $\delta y$ of the string.
  $X(t)=|x_{N}(t)-x_{0}(t)|$. $\delta y = \sqrt{\sum_{i=0}^{N}y_i^2/N}$. $t$ is
  measured in the unit of $200\tau_0$. $A_{\textrm {noise}}=10^{-5}\ell_0$.
  $a/a_0=10^{-3}$.  $N=50$. 
  }
\label{wiggling}
\end{figure}

In fig.~\ref{acc}, we present typical snapshots of wiggling strings. $y_i$ is
the transverse displacement of each bead. The head bead is labeled as $i=0$. The
magnitude of acceleration increases from the green to the red lines.  From
fig.~\ref{acc}, we see that the tail of the string is generally subject to a
stronger shape fluctuation than the head part. Increasing the acceleration
enhances the strength of string wiggling. Wiggling transition still
occurs by reducing the noise level to as low as $A_{\textrm
{noise}}=10^{-5}\ell_0$.

Considering that the string in simulations is not strictly inextensible, is it
possible that the wiggling of the string is caused by the extensibility of the
string?  To clarify this question, we perform theoretical analysis based on
eqs.(\ref{eom1}) and (\ref{eom2}) for inextensible strings. Furthermore,
theoretical analysis based on the string equations allows us to
explore the inextensible
regime which is beyond the applicability of our numerical simulations. Here, we emphasize
that our numerical simulations are based on the spring-bead
model with large spring constant (but not strictly inextensible), and the string equations are
for inextensible strings. 

We focus on the behavior of
the string at the onset of wiggling transition when the shape fluctuation is
small and varies slowly over the string. This justifies a continuum description
of the string based on the equations of motion. 
 The main results are presented below. The stress
distribution can be
written as $\sigma(s, t) = f(t) s + \alpha(t)$. The requirement of a stress-free
end at $s=0$ sets $\alpha$ to be zero. The shape of
the string is represented by the orientation of the tangent vector $\theta$ with
respect to the x-axis. $\theta(s, t) = \theta_1(s) \theta_2(t)$,
where $\theta_1(s) =\theta_{10} \exp(-qs)$, and $\theta_2(t)$ satisfies
  \begin{eqnarray}
\ddot{\theta}_2(t)+qg(t)\theta_2(t)=0, \label{theta2}
\end{eqnarray}
where $g(t)=2f(t)/\mu$ and $q$ is a constant. 
It is of interest to note that eq.(\ref{theta2}) has the same mathematical form
as the Schr$\ddot{\textrm o}$dinger equation; $g(t)$, the time-dependent part of the stress
$\sigma(s, t)$,
corresponds to the physical potential in the Schr$\ddot{\textrm o}$dinger equation.
Equation (\ref{theta2}) suggests the rich dynamics of the string even in the 
perturbation regime.

\begin{figure}[t]  
\centering 
\includegraphics[width=1.7in]{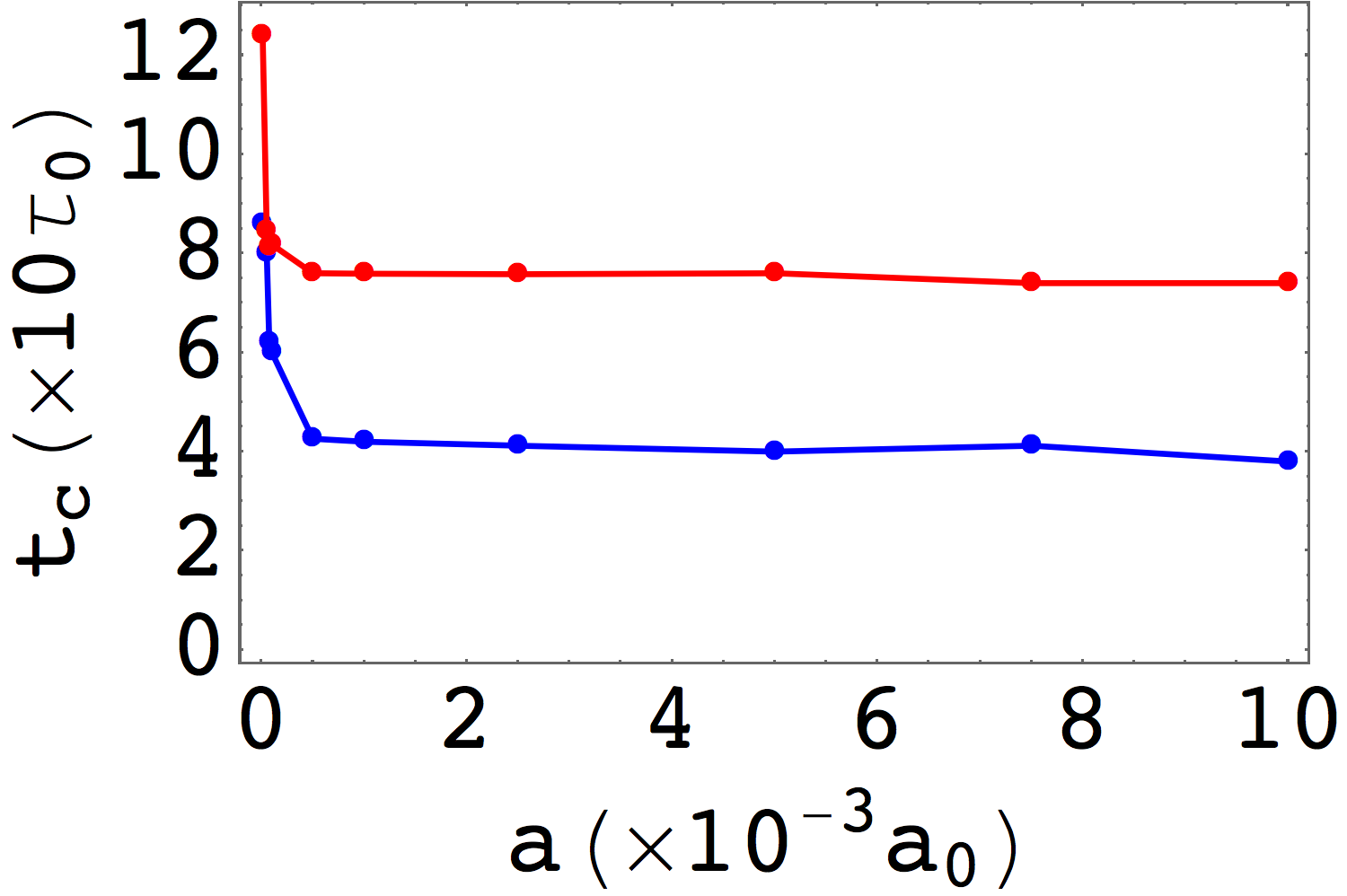}
\caption{Phase diagram of the string in uniform acceleration. The curves of
  $N=50$ (bottom, blue) and $N=100$ (top, red) indicate the transition of the
  dynamic state of the string from the straight to the wiggling state.
  $A_{\textrm {noise}}=10^{-5}\ell_0$.
   }
\label{wiggling_pt}
\end{figure}

Now, consider the case of interest: $g(t)=g_0$. $g_0$ is a constant, and $g_0>0$
without loss of generality. Such a distribution of stress is identical to that
over a straight string in uniform acceleration $a=g_0/2$. By inserting
$\theta_2(t) = \theta_{20}\exp(iwt)$ into eq.(\ref{theta2}), we obtain the
dispersion relation: $(iw)^2 = -qg_0$. For real positive $q$, $w = \sqrt{qg_0}$.
Such a solution is plotted in fig.~\ref{acc}(b). The tail of the string wiggles,
and the spatial extension $q^{-1}$ of the wiggling segment is linear with the
magnitude of acceleration at fixed frequency $w$. Therefore, in addition to the
trivial straight-string solution, the tail of an uniformly accelerating,
inextensible string can wiggle.  This analytical result
and the preceding simulation results suggest that the extensibility of the string is not a
necessary condition for the occurrence of the string wiggling, but it may
contribute to the propagation of the wiggling deformation  to the entire string.
Here, it is of interest to point out that the
wiggling transition is an intrinsic property of the string without
dependence on any external transverse force.

In the following, we further characterize the wiggling transition by the
evolution of its longitudinal length $X$ and the 
averaged transverse displacement $\delta y$.  $\delta y(t) = \sqrt{\sum_{i=0}^{N}y_i^2(t)/N}$.
Figure~\ref{wiggling} shows that the transition from the straight to the
wiggling state is well signified by the entire decline of the oscillations in
$X$, and the simultaneously occurring take-off of the $\delta y(t)$ curve from
the zero line. Long-time observation up to $t=10$ millions simulation steps
shows the convergence of the string wiggling; the strength of wiggling remains
in the interval of $\delta y \in [0.12, 0.20]$, and $X \in [50.2, 51.0]$.

In fig.~\ref{wiggling_pt}, we present the phase diagram of the dynamic state of
the string under longitudinal uniform acceleration. The lower (blue) and upper
(red) curves are for the cases of $N=50$ and $N=100$, respectively.  The state
of the string is characterized by the averaged transverse displacement $\delta
y$. The string is regarded to be in the wiggling state when $\delta y$ exceeds
ten times the initially introduced noise $A_{\textrm {noise}}$.
Figure~\ref{wiggling_pt} shows that a longer string can stay in the straight state
for a longer time. The straight-to-wiggling transition becomes insensitive to the
magnitude of acceleration when it exceeds about $10^{-3}a_0$.

\section{Conclusions}

To summarize, in this work we investigated the planar dynamics of a flexible
string that is subject to transverse and longitudinal motions at one end. We
revealed the pulse structure in the propagation of
stress when one end of the string is under transverse harmonic motion, and
identified the wiggling transition in a traveling string in uniform acceleration.
These results may find applications in the remote control of various
filamentary thin structures by manipulating the end.

\acknowledgments

This work was supported by NSFC Grant No. 16Z103010253, the SJTU startup fund
under Grant No. WF220441904, and the award of the Chinese Thousand Talents
Program for Distinguished Young Scholars under Grant No.16Z127060004 and
No. 17Z127060032. The author thanks Kai Yao for stimulating discussions.

\end{document}